\documentclass[fleqn,usenatbib]{mnras}

\usepackage{newtxtext,newtxmath}  

\usepackage[T1]{fontenc}

\usepackage{graphicx}  

\usepackage{amssymb}    
\usepackage{amsmath}    
\usepackage{float}      

\def\hi{H\,{\sc i}}

\def\msun{$M_{\odot}$}

\def\msunpc{M$_{\odot}$~pc$^{-2}$}

\def\simba{\textsc{Simba}}

\title[The \simba\ \hi\ mass-size relation]{Investigating the \hi\ mass-size relation using the \simba\ cosmological simulations.}

\author[Omphile Rabyang]{
Omphile Rabyang,$^{1}$\thanks{E-mail: 4115981@myuwc.ac.za}
Ed Elson,$^{1}$
\\
$^{1}$Department of Physics $\&$ Astronomy, University of the Western Cape, Robert Sobukwe Rd, Bellville, 7535, South Africa
}

\date{Accepted XXX. Received YYY; in original form ZZZ}

\pubyear{\the\year{}}

\begin{document}
\label{firstpage}
\pagerange{\pageref{firstpage}--\pageref{lastpage}}
\maketitle

\begin{abstract}
Observational studies have established a remarkably tight power-law relationship between the \hi\ masses and sizes of late-type galaxies, known as the \hi\ mass–size relation. This relation has been shown to persist across various models of a galaxy’s \hi\ surface density profile. Using the \simba\ cosmological simulations, we investigate the robustness of this relation under different feedback prescriptions, including cases where specific feedback mechanisms are absent. While the global properties of galaxies are significantly affected by changes in feedback, the \hi\ mass–size relation remains intact. Moreover, its parameters consistently align with the best available empirical measurements. We analyse the \hi\ mass distributions of galaxies and demonstrate that, regardless of the feedback scenario, galaxies within a given \hi\ mass bin exhibit outer \hi\ radial profiles well-approximated by an exponential function. Furthermore, the exponential decline rate remains remarkably similar across different physical prescriptions.  This behaviour is only valid within the  $\sim$ 1 \msunpc\ contour. We attribute the persistence of the \hi\ mass–size relation to this inherent self-similarity in the \hi\ mass distributions.  
\end{abstract}




\section{Introduction}

The spatial distribution of neutral atomic hydrogen (\hi) in galaxies provides crucial insights into their assembly histories, mechanisms of gas accretion and the role of feedback processes in shaping their evolution. Among the most robust observational findings in this context is the tight empirical relation between a galaxy’s \hi\ mass and its spatial extent, known as the \hi\ mass–size relation (\hi MSR) \citep{broeils1997short,wang2016h}. This relation is well described by a single power law over several orders of magnitude in \hi\ mass and is characterized by remarkably low scatter, typically around 0.06 dex, indicating a high degree of regularity across diverse galaxy populations.\\

The universality and tightness of the \hi MSR are widely interpreted as evidence for structural self-similarity in \hi\ discs, suggesting that similar physical mechanisms govern the distribution of atomic hydrogen across a broad range of galaxy types and environments \citep{wang2014observational,stevens2019origin}. Observational studies have found that the average \hi\ surface density, $\Sigma_{\text{\hi}}(r)$, within the \hi\ radius is nearly constant for most galaxies, further supporting the idea of a common underlying process. 
\\

Despite this, the extent to which feedback processes— arising from both stellar processes and active galactic nuclei (AGN) activity—might disrupt or regulate the \hi MSR remains an open question. Theoretical and semi-analytic models suggest that feedback can influence the structure of \hi\ discs, but the observed tightness of the \hi MSR places strong constraints on the degree of disruption permissible. Recent hydrodynamical simulations, such as EAGLE \citep{crain2015eagle,bahe2016distribution}, have shown that strong AGN feedback can suppress central \hi\ densities and potentially induce deviations from the canonical \hi MSR, highlighting the need to isolate the influence of individual feedback channels. \\

In this study, we investigate the impact of various feedback prescriptions on the \hi MSR using the \simba\ \citep{dave2019simba} suite of cosmological hydrodynamical simulations. \simba\ incorporates both stellar and AGN feedback, including radiative, jet, and X-ray heating modes, enabling a systematic exploration of their respective roles. We analyse four variants of the simulation: a fiducial run with full feedback physics, and three others in which one or more AGN feedback modes are selectively disabled. For each scenario, we measure the global \hi MSR and examine the radial \hi\ surface density profiles to assess whether the observed universality arises from structural self-similarity and whether this is preserved under different feedback regimes.\\

This paper is structured as follows. In Section \ref{SIMBA_description}, we describe the simulation and the different variants. Section \ref{sec:ellipse_fitting} focuses on the  methodology. Section \ref{sample} presents the sample selection for the \hi MSR. The results and  discussion are presented in  Section \ref{sec:results} and \ref{sec:discussion}, respectively.

\section{\simba\ Simulations} \label{SIMBA_description}
In this study, we use the \simba\ suite of cosmological simulations \citep{dave2019simba}, a descendant of the MUFASA simulation \citep{dave2016mufasa}, which is built upon the meshless finite mass (MFM) hydrodynamics solver of the GIZMO code \citep{hopkins2015new}. \simba\  incorporates updated physics for black hole growth and feedback and includes prescriptions for star formation, black hole seeding and accretion, as well as stellar and AGN feedback. It also accounts for radiative cooling, photoionization heating, metal cooling, and the out-of-equilibrium evolution of primordial elements via the grackle-3.1 library \citep{smith2017grackle}. \simba\  calculates molecular hydrogen content on-the-fly and uses a subgrid model for star formation, which allows it to account for the molecular phase of gas.

The simulation models black hole accretion using both Bondi-Hoyle accretion for hot gas and torque-limited accretion for cold gas \citep{angles2017black}, improving realism compared to other simulations that use only Bondi accretion. There are two main types of feedback in \simba: stellar feedback and AGN feedback. Stellar feedback incorporates supernovae, radiation pressure, and stellar winds, while AGN feedback operates in three modes \footnote{X-ray feedback injects thermal energy  into diffuse gas and a mix of kinetic and thermal energy into dense ISM gas}:

\begin{enumerate} 
\item \textbf{AGN Winds:} Active when the Eddington ratio is high (f$_{\text{Ed}} >$ 0.2). In this regime, black holes (BH) drive mass-loaded winds with velocities of $\sim$ 1000 km$\cdot$s$^{-1}$, ejected radially in the direction of the angular momentum of the inner accretion disk. These outflows are launched with zero opening angle and do not alter the gas temperature at injection.\\
\item \textbf{AGN Jets:} Fully activated at low Eddington ratios (f$_{\text{Ed}} <$ 0.02) for BH with masses $> 10^{7.5}$ M$_{\odot}$. This mode produces highly collimated, bipolar jets with zero initial aperture and velocities reaching up to 7000 km$\cdot$s$^{-1}$. The outflowing gas is heated to the halo virial temperature prior to ejection. In the intermediate regime (0.02$<$f$_{\text{Ed}}<$0.2), wind velocity increases smoothly as f$_{\text{Ed}} $ decreases.\\
\item \textbf{X-ray Heating:}  Activated concurrently with jet mode. Based on the model of \citet{choi2012radiative}, this mode differentiates its treatment of gas by density. For non-interstellar medium (ISM) gas, the injected X-ray energy directly increases the temperature. For ISM  gas, half the energy is used to impart a radial velocity kick, while the remainder is deposited as heat. This feedback component is important for suppressing residual star formation in quenched galaxies.
\end{enumerate}


The AGN feedback implementations in   \simba\ are designed to reproduce the observed bimodality in black hole growth in AGN, as reported by \cite{heckman2014coevolution}. \simba\  also implements a two-mode kinetic AGN feedback, with radiative winds at high accretion rates and jet feedback at lower accretion rates. The combination of these feedback mechanisms helps to quench massive galaxies and match observed galaxy populations and BH growth. Finally, \simba\  models the transformation of ionised gas into atomic and molecular phases by incorporating self-shielding and using a prescription for the H$_{2}$ fraction based on the gas metallicity, as described in the \citet{rahmati2013evolution}.
Since \simba\  has been extensively described in previous works, we provide a brief overview of the main simulation parameters and refer the reader to \citet{dave2019simba} for a more detailed description. Additional information on the feedback mechanism modes of \simba\  can also be found in \citet{ward2022cosmological,christiansen2020jet,khrykin2024cosmic}.\\

In these simulations, halos are identified on the fly using a 3-D friends-of-friends (FOF) algorithm integrated into Gizmo, based on Gadget-3, with a linking length of 0.2 times the mean inter-particle separation. The YT-based package CAESAR\footnote{\href{https://caesar.readthedocs.io/en/latest}{https://caesar.readthedocs.io/en/latest}} is then used in post-processing to cross-match galaxies and halos and generate a catalogue of pre-computed properties. Additionally, galaxies are identified using a 6-D FOF galaxy finder with a spatial linking length of 0.0056 times the mean inter-particle separation (twice the minimum softening kernel) and a velocity linking length set to the local velocity dispersion. This is applied to all stars and ISM gas with $n_{\text{H}}$ > 0.13 cm$^{-2}$. The \hi\ and H$_{2}$ fractions for individual gas elements are taken directly from the simulation, without post-processing.
To compute the \hi\ and H$_{2}$ contents of galaxies, each gas particle in a halo is assigned to the galaxy with the highest $M_{\text{baryon}}$/$R^{2}$ ratio, where $M_{\text{baryon}}$ is the total baryonic mass and $R$ is the distance from the particle to the galaxy’s centre of mass. This ensures that cold gas, particularly \hi\, can be assigned to a galaxy even if it is not part of its  ISM, as \hi\ can be significant even for gas with $n_{\text{H}}$ > 0.13 cm$^{-2}$. It is important to note that black holes and \hi\ gas are assigned to the galaxy to which they are most gravitationally bound, with the most massive black hole particles being designated as the central black hole.

\subsection{\simba\ feedback variants}
This study utilises five variants of the \simba\ suite of hydrodynamic simulations, all initialised with identical conditions. Each simulation box spans a co-moving volume of 50 h$^{-1}$Mpc per side and contains an equal number of dark matter and gas particles, specifically 512$^{3}$ for each type, ensuring uniform mass resolution of 9.6$\times$ 10$^{7}$ \msun\ for the dark matter and 1.82$\times$ 10$^{7}$ \msun\  the gas particles.  This corresponds to an \hi\ mass resolution of  $\sim 10^{8}$  \msun. The cosmological framework for these simulations aligns with the Planck-16 $\Lambda$CDM model, characterised by parameters: $\Omega_{m}$ = 0.3, $\Omega_{\Lambda}$ = 0.7, $\Omega_{b}$ = 0.048, h = 0.68, $\sigma_{8}$ = 0.82 and n$_{s}$ = 0.97. The s50 simulation serves as the fiducial run, incorporating all features outlined in Section (\ref{SIMBA_description}). The other four runs progressively deactivate the feedback modules, as summarised in Table \ref{tab:stats_FB}. The  s50noX run excludes only the X-ray mode, while the s50nojet run deactivates both jets and X-ray heating. In the s50noAGN run, all AGN feedback modes are turned off, leaving only stellar feedback processes active. Lastly, the s50nofb run, all feedback mechanisms- including both stellar and AGN feedback- are turned off. It is important to note that \simba\ still spawns black holes and includes metal cooling. Clarify that only stellar and AGN feedback is removed, not cooling, enrichment or H$_{2}$ formation. 
It is important to note that both the s50nofb and s50noAGN runs still include black hole seeds and the dual accretion mode. Due to the self-regulation of the torque-limited accretion mode (as described in \citet{angles2017black,dave2019simba}), black hole-galaxy scaling relations remain intact, even when combined with Bondi accretion and without feedback. For this study, only the $z = 0$ snapshot  from each of the five runs is used. \\




\begin{table}
\centering
\caption{Summary of feedback prescriptions used in the simulation runs. Each column represents a specific feedback mechanism—stellar feedback, AGN winds, AGN jets, and X-ray heating. A check mark (\checkmark) indicates that the corresponding mechanism is included in a given run, while a dash (–) denotes its absence. Rows correspond to individual simulation runs, as described in the manuscript.
}

\begin{tabular}{ |p{1.5cm}||p{0.8cm}|p{0.8cm}|p{0.8cm}|p{2cm}| }
 \hline
 \multicolumn{5}{|c|}{                Feedback Prescriptions} \\
 \hline
 Name & Stellar & AGN Winds & AGN Jet & X-ray Heating \\
 \hline
 s50 & \checkmark & \checkmark & \checkmark & \checkmark \\
 s50noX & \checkmark & \checkmark & \checkmark & - \\
 s50nojet & \checkmark & \checkmark & - & - \\
 s50noAGN & \checkmark & - & - & - \\
 s50nofb & - & - & - & - \\
 \hline
\end{tabular}
\label{tab:stats_FB}
\end{table}

\section{Methods}\label{sec:ellipse_fitting}

In this work, we adopt an approach that provides a more observationally consistent method for measuring the \hi\ masses and sizes of galaxies, ensuring that our results are more directly comparable to empirical studies by minimising potential biases arising from differences in measurement techniques. While directly using \simba\ particle lists offers a theoretically precise measure, it arguably lacks the observational realism necessary for a meaningful comparison.  

\begin{figure*}
    \includegraphics[scale=0.53]{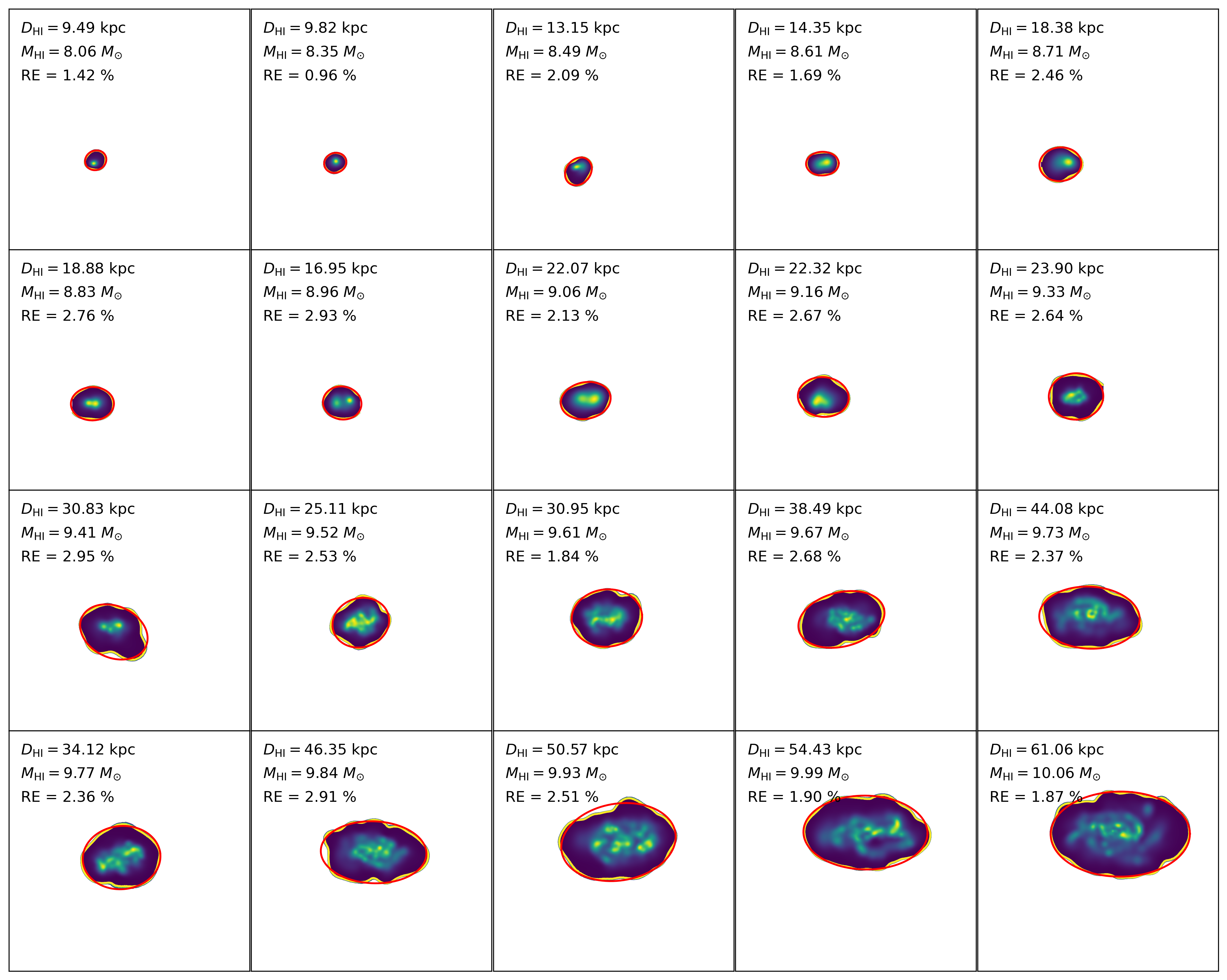}
    \caption{Twenty selected \hi\ total intensity maps of galaxies from the full-physics (s50) run. The \hi\ mass increases from left to right and top to bottom. The yellow contour in each panel marks the 1~\msunpc\ surface density level, while the fitted ellipse is shown in red. The text in each panel provides the physical size of the fitted ellipse, the galaxy’s total \hi\ mass, and the relative error percentage , $\Delta$ D$_{\text{\hi}}$/D$_{\text{\hi}}$ , between the fitted ellipse and the 1~\msunpc\ density contour. }
    \label{fig:mom0_panel}
\end{figure*}

For each galaxy in our samples, we use the \textsc{Martini} (Mock APERTIF-like Radio Telescope Interferometry of the Neutral ISM) package \citep{martini_1, martini_2, martini_3} to generate a synthetic \hi\ line data cube and derive an \hi\ total intensity map. All cubes have a spatial pixel size of 8 arcseconds, and a Gaussian point-spread function with a half-power width of 24 arcseconds is applied for spatial smoothing. A fixed distance of 4 Mpc is assumed for all galaxies. The \hi\ disc of each galaxy is inclined at 60 degrees to the y-axis in all cubes. Martini attempts to determine a preferred disc plane for each galaxy based on the angular momenta of the central one-third of its particles. Once an \hi\ data cube is generated, its flux is spectrally integrated to produce a total intensity map.  After converting the map’s pixel units from Jy/beam to \msunpc, we identify a thin mass surface density contour centered on $\sim$ 1~\msunpc\ and fit it with an ellipse. The major axis of the fitted ellipse defines the galaxy’s \hi\ diameter, while the \hi\ mass is obtained by summing the flux of all pixels with a surface density $\ge 1$ \msunpc.  To identify galaxies with outer \hi\ distributions too disturbed to yield reliable size estimates, we compute a quality-of-fit parameter for each fitted ellipse. For every point along the ellipse, we calculate its distance to the nearest map pixel\footnote{Within the thin mass surface density contour centred on 1~\msunpc.}. The mean of these distances is then normalised by the semi-major axis of the fitted ellipse, providing a relative measure of fit quality. In Section~\ref{sample}, we describe how this quality parameter is used to refine our galaxy samples.  Fig. \ref{fig:mom0_panel} presents the \hi\ total intensity maps and fitted ellipses for 20 galaxies that uniformly span the mass range $8 \leq \log_{10}\left(\frac{M_{\text{HI}}}{M_\odot}\right) \leq 10$ from the full-physics (s50) run, ordered by increasing \hi\ mass.  

\section{Sample} \label{sample}

To create a suitable sample, we use the CAESAR-computed galaxy properties to select a subset best suited for measuring the quantities required to construct the \hi MSR. Previous studies such as \citet{glowacki2020baryonic,glowacki2021redshift,elson2023measurements} applied stellar and \hi\ mass cuts of M$_{*}$ $>$ 7.25$\times$10$^{8}$ \msun\ and M$_{\text{\hi}}$ $>$ 1.25$\times$10$^{8}$ \msun, respectively. These cuts ensure that galaxies from the 25 Mpc$^{3}$ \simba\ run are well above the mass resolution limit, making their measured properties accurate and reliable. In this work, we apply the same mass cuts to galaxies in the 50 Mpc$^{3}$ runs, even though their typical mass resolutions are expected to be significantly higher. This allows us to explore how far below the expected resolution limit our measured \hi MSR extends. However, when parametrising the relation, we adopt more appropriate mass limits. Further details are provided in Section ~(\ref{sec:results}).\\

To minimise the impact of galaxy–galaxy interactions on our \hi\ measurements, we exclude all galaxies with at least one neighbour within $\sim$30 kpc. Additionally, to ensure our sample consists primarily of galaxies with extended, regular, rotating \hi\ discs, we impose a dynamical morphology cut. Following \citet{sales2012origin}, we define:

\begin{eqnarray} \kappa_{\rm rot} = \frac{K_{\rm rot}}{K} ,\end{eqnarray}

where $K_{\rm rot}$ is the kinetic energy in ordered rotation and $K$ is the total kinetic energy of the galaxy. We retain only galaxies with $\kappa_{\rm rot} > 0.8$, as measured for their \hi\ component.\\

As described in Section~(\ref{sec:ellipse_fitting}), each galaxy is fitted with an ellipse along a thin mass surface density contour centered on  $\sim$1 \msunpc. We retain only those galaxies for which the mean deviation between the ellipse and the contour is less than 5\% of the semi-major axis length. This selection criterion effectively removes galaxies with significantly disturbed outer \hi\ discs.\\

\begin{figure*}
    \centering
    \includegraphics[scale =0.5
    ]{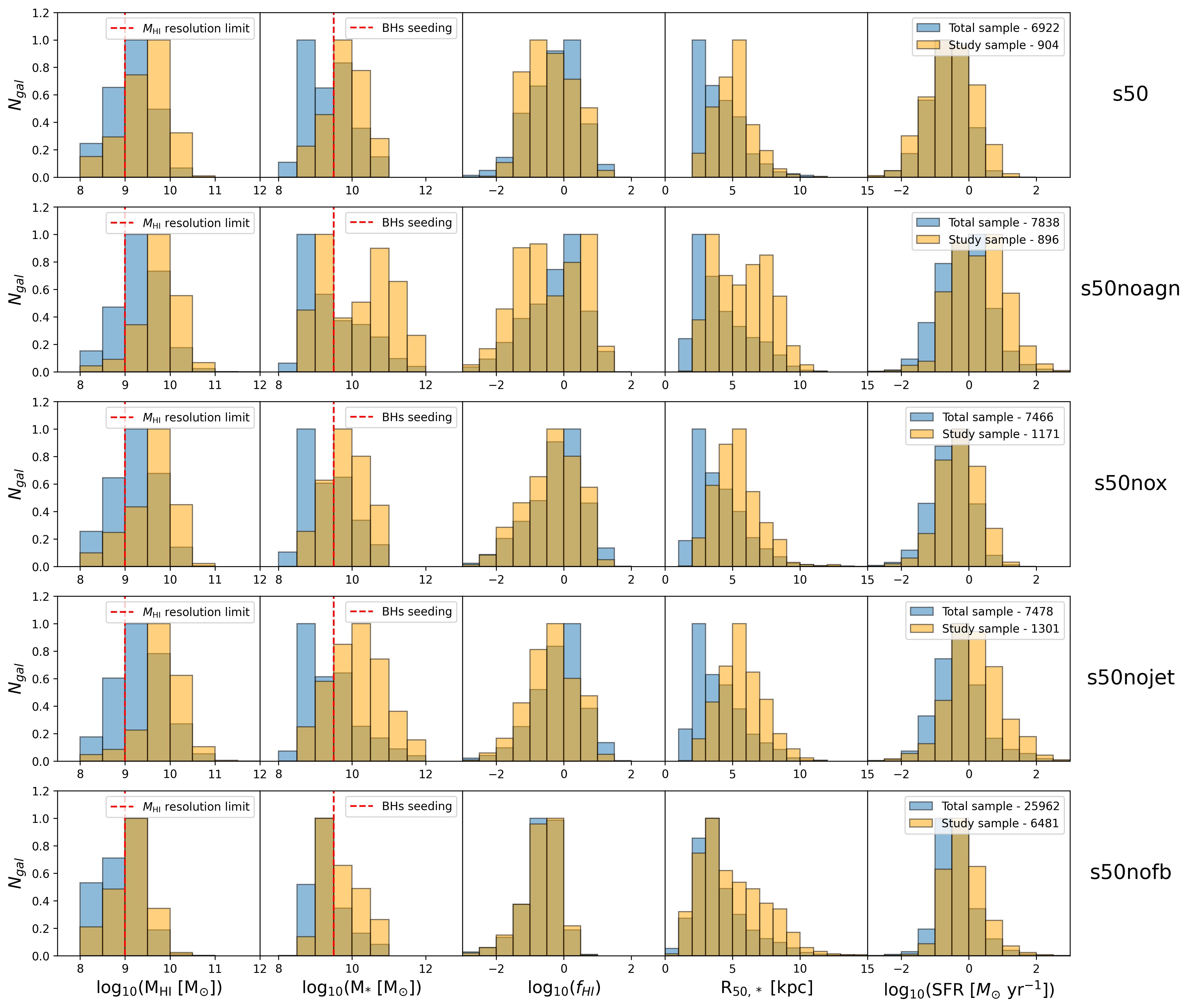}
    \caption{Distributions of key galaxy properties—logarithmic \hi\ mass, stellar mass, \hi\ fraction, and stellar half-mass radius—are shown for five simulation variants: s50, s50noAGN, s50noX, s50nojet, and s50nofb. In each panel, histograms compare the full galaxy sample (blue) with the selected study sample (orange). The variant name is indicated to the right of each row. The fourth panel also includes the number of galaxies in each sample. Vertical red dashed lines denote resolution limits: M$_{\text{\hi}} >$ 10$^{9.1}$ \msun\ for \hi\ mass and M$_{*} >$ 10$^{9.5}$ \msun, the threshold in \simba\ where black holes are seeded once a galaxy reaches this stellar mass limit, the latter corresponding to the stellar mass threshold in Simba above which black holes are seeded. Bin sizes are 0.5 dex for all properties except R$_{50,*}$ which uses 1 dex bins.
}
    \label{fig:sample_selection}
\end{figure*}

For each of the 50 Mpc$^{3}$ runs used in this study, Fig.~\ref{fig:sample_selection} shows the distribution of various galaxy properties. The blue histograms represent the full galaxy sample, while the orange histograms correspond to the subset used to generate the \hi\ MSR. \\



In the s50 run, the stellar and \hi\ mass distributions are relatively narrow and centered around $\sim$10$^{9-10}$ \msun, reflecting a self-regulated galaxy population. In contrast, runs lacking one or more key feedback prescriptions—such as AGN winds, jets, or X-ray feedback—exhibit broader distributions with extended tails toward both higher and lower masses. For example, the s50nofb variant shows a clear shift to lower \hi\ masses, as the absence of feedback allows for excessive cooling. Gas fractions also vary significantly: runs with full feedback maintain moderate, balanced gas reservoirs, while feedback-deficient runs display extreme values due to either rapid gas consumption or inefficient gas retention. Additionally, galaxies in the no-feedback variants tend to be more compact and show a wider spread in star formation rates—from bursty to quenched—compared to the more uniform 2–5 kpc half-light radius and moderate star formation rates seen in the full-physics sample. All of these trends exemplify the expected impact of missing feedback: without proper regulation, galaxies can form across a wider range of masses and exhibit more diverse gas content, sizes, and star formation histories. The main aim of this study is to determine the resilience of the \hi\ MSR to these differences.

\section{Results} \label{sec:results}

\begin{figure*}
    \centering
    \includegraphics[scale=0.42]{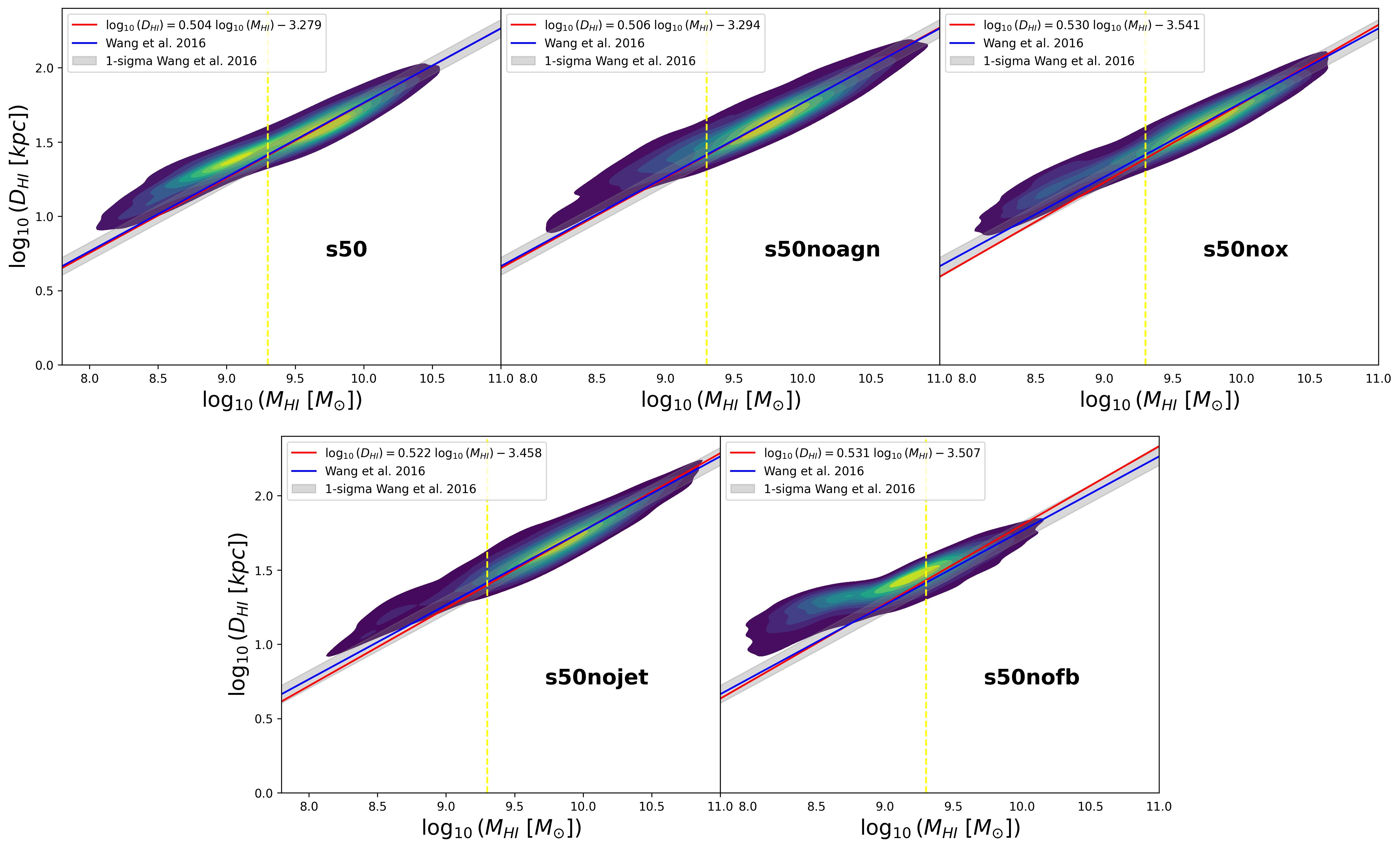}
   
   \caption{This plot represents 5 different \hi\  mass-size relations (\hi MSR) from the \simba-50 feedback variants. Each subplot shows the kernel density estimate (KDE) of the data with a colour map, along with the linear best-fit line (in red) derived from orthogonal distance regression (ODR) algorithm. The blue line represents the relation from \citet{wang2016h}, with the grey shaded region indicating the 1$\sigma$ uncertainty of their fit. A vertical dashed line at log$_{10}$(M$_{\text{\hi}}$[\msun]) = 9.1 marks the threshold at which the best-fit line is fitted.}
    \label{fig:All_HIMSR}
\end{figure*}

Fig. (\ref{fig:All_HIMSR}) presents the fitted \hi\ diameters as a function of  \hi\ mass for the five simulation variants. Galaxies in the \simba\ 50 Mpc/h boxes simulations are well resolved at $\log_{10}$(M$_{\text{\hi}}$[\msun]) $\ge$ 9.  The resolution limit for the 25 Mpc/h box is approximately 1.25$\times$ 10$^{8}$M$_{\odot}$, which is eight times  finer than that of the 50 Mpc/h box, where the corresponding limit is therefore expected to be close to 1$\times$ 10$^{9}$M$_{\odot}$. To ensure a reliable fit to the \hi SMR, we adopt a conservative threshold of log$_{10}$(M$_{\text{\hi}}$[\msun]) $\geq$ 9.1, providing a small, additional safety margin above the resolution limit. This  criterion increases the chances that galaxies included in the \hi SMR  are well resolved (numerically), minimising biases from under-resolved systems and enabling a robust characterisation of the relation across the full sample. 
\\

Overlaid on each panel is the best-fitting power-law model for our fits.





\begin{eqnarray}
   \log_{10} \left( \frac{D_{\text{\hi}}}{\text{kpc}} \right) = \alpha \cdot \log_{10} \left( \frac{M_{\text{\hi}}}{M_\odot} \right) - \beta ,
   \label{eqn:2}
\end{eqnarray}

Obtained via Orthogonal Distance Regression (ODR; \citealt{brown1990statistical}), which minimizes perpendicular distance by accounting for errors in both variables. We quantify the model's performance using the root-mean-square perpendicular distance (for galaxies above the 10$^{9.1}$ \msun threshold. Table (\ref{tab:stats}) summarized the best-fitting parameters and the associated scatters.)  \\

Also shown in each panel (blue solid line) is the empirical \hi MSR from \cite{wang2016h}.  When fitting Eqn~(\ref{eqn:2}) to all galaxies with $\log_{10}\left( \frac{M_{\mathrm{HI}}}{M_\odot} \right) \ge 9.1$. All simulation runs agree well with the empirical relation despite the absence of certain feedback prescriptions. In the following section, we discuss why these relations persist.

\begin{table*}
\caption{\hi MSR  fits to our sample of 50 Mpc$^{3}$ box for the different feedback variants. }
\begin{tabular}[t]{lcccccc}
\hline
&Wang et al 2016&s50&s50nox&s50nojet&s50noagn&s50nofb\\
\hline
Sample & 562&904&1171&1301&896&6481\\
     $\alpha$&0.506 $\pm$ 0.003&0.504 $\pm$ 0.010&0.530 $\pm$ 0.008&0.522 $\pm$ 0.006&0.506 $\pm$ 0.009&0.531 $\pm$ 0.006\\
     $\beta$&-3.293 $\pm$ 0.009&-3.279 $\pm$ 0.093&-3.541 $\pm$ 0.075&-3.458 $\pm$ 0.059&-3.294$\pm$ 0.085& -3.507 $\pm$ 0.056\\
     $\sigma$ [dex]&0.06&0.096&0.103& 0.083&0.097&0.122\\
     \hline
\end{tabular}
\label{tab:stats}

\end{table*}
\section{Discussion}\label{sec:discussion}

The \hi MSR has been shown to be well described by a single power-law spanning a wide range of \hi\ masses (e.g. \citealt{wang2016h}). This apparent universality is often interpreted as evidence for the self-similarity in the spatial distribution of gas within galaxies, where similar physical processes govern both the acquisition and distribution of \hi\, irrespective of galaxy size or mass. Analytical work by \citet{stevens2019origin} demonstrates that when \hi\ mass distributions exhibit self-similarity—particularly in their outer radial regions—a natural scaling arises between total \hi\ content and spatial extent. Observational studies by  \cite{broeils1997short, verheijen2001ursa, wang2016h} and \cite{naluminsa2021h}  further support this interpretation, showing that the tightness of the relation is likely underpinned by the structural uniformity of \hi\ discs.\\

\begin{figure*}
\centering
    \includegraphics[scale = 0.35, trim=0 0 0 100, clip]{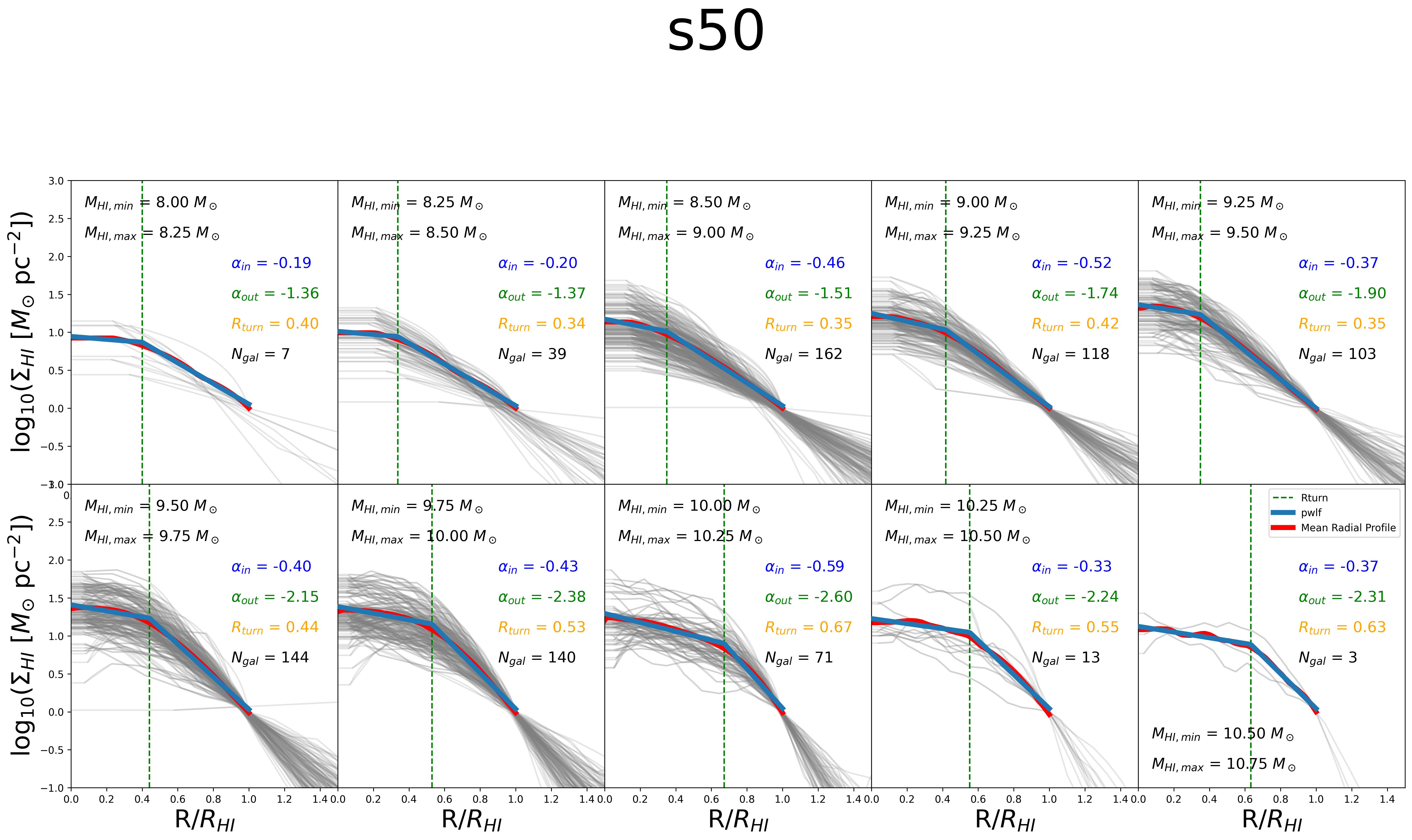}
   \caption{The scaled \hi\ radial profiles from the \simba-50 z = 0 are shown. These profiles are scaled to $R_{\text{\hi}}$, and by definition, they overlap at a surface density of  $\Sigma_{\text{\hi}} =$ 1 M$_{\odot}$.pc$^{-2}$ . Each panel corresponds to a  different mass bins, with a width of 0.25 in log-scale, covering the range log$_{10}$($M_{\text{\hi}}$) = [8,10.75]. The number of galaxies per mass bin is denoted as N$_{gal}$. The grey lines represent the individual galaxy profiles, the red line shows the mean \hi\ radial profile. The green dashed line indicates $R_{\mathrm{turn}}$ , the turning point that separates the inner and outer regions of the \hi\ profiles. The blue line  represents the fit to the mean \hi\ radial profile, obtained using the piecewise linear fitting (\textsc{pwlf}) method in Python. $\alpha_{in}$ and $\alpha_{out}$ represents the inner and outer slopes.}
    \label{fig:HI-radial-profile-s50}
\end{figure*}

Here, we investigate whether a similar self-similarity holds for the different samples of \simba\ galaxies, all of which have been shown to yield similar \hi MSRs. For each galaxy in a given sample, the orientation parameters of its best-fitting ellipse \footnote{Used to measure its diameter for the \hi MSR.} are used to divide its \hi\ total intensity map into concentric elliptical rings. The \hi\ flux in each ring is azimuthally averaged to generate a radial mass profile. Each profile is then normalised in radius by the galaxy’s measured, $R_\text{\hi}$. Galaxies are grouped into \hi\ mass bins of width 0.25 dex over the range $\log(M_\text{\hi} / M_\odot) = $ 8–$10.75$, and within each bin, the radial profiles are averaged.\\

\begin{figure*}
    \includegraphics[scale = 0.35]{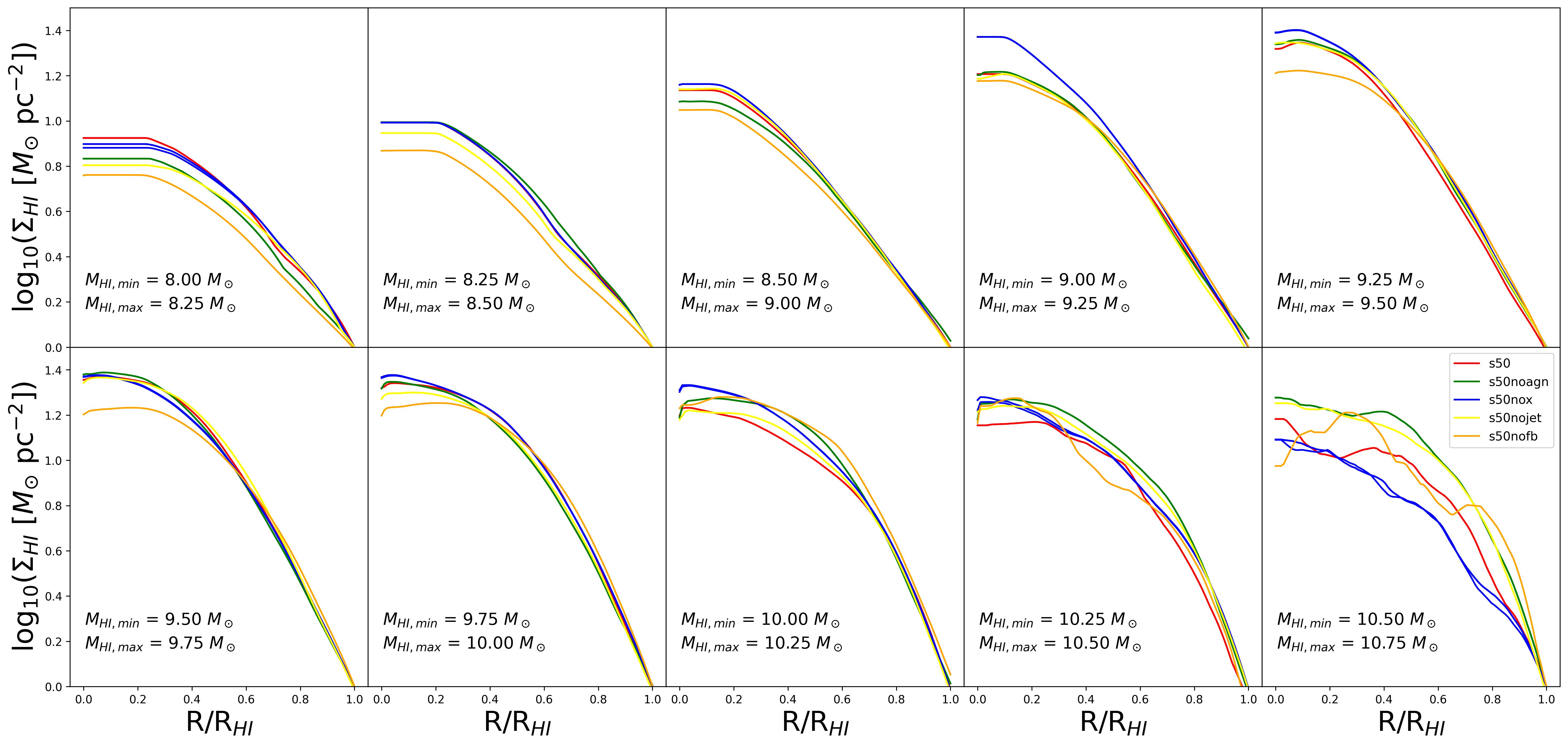}
    \caption{The scaled mean \hi\ radial profiles from the \simba-50 simulation at z=0 for different feedback variants. When normalised by each galaxy’s characteristic \hi\ disc size, the profiles reveal a universal exponential shape across all \hi\ mass bins and feedback scenarios, highlighting the self-similarity of the outer \hi\ distributions regardless of feedback strength.}
    \label{fig:All_HI_radial_profiles}
\end{figure*}


From left to right and top to bottom, the panels in Fig.~\ref{fig:HI-radial-profile-s50},  show the \hi\ radial profiles normalised in radius from the various \simba\ 50 Mpc runs.  In each panel, grey curves represent the profiles of individual galaxies in the give mass bin, while the red curve represents the mean profile. In this study we model each mean \hi\ radial profile as a piecewise linear function comprising two segments—representing the inner and outer parts —naturally capturing the turnover radius, $R_{\mathrm{turn}}$.  This is modelled using the \textsc{pwlf} Python package, with the resulting fits shown as blue lines in Fig.~\ref{fig:HI-radial-profile-s50} the corresponding turnover radius, $R_{\mathrm{turn}}$, is indicated by the dashed green line.\\

\begin{figure}
    \centering
    \includegraphics[scale = 0.5]{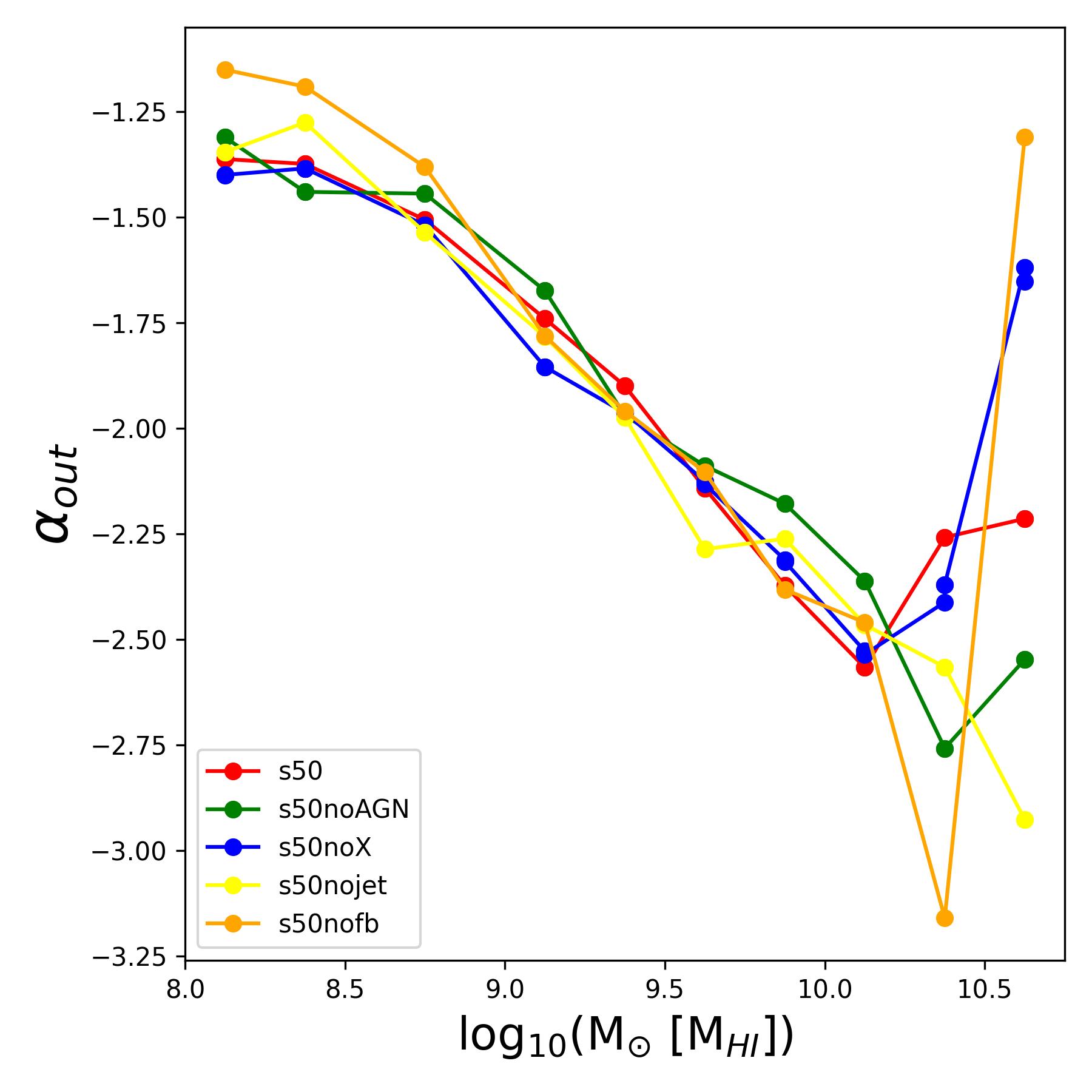}
    \caption{The outer slope as a functions of the mass bin midpoint for different simulation variants. This figure illustrates that the outer slope steepens with increasing mass, with feedback mechanisms influencing the rate of decline, but the behaviour is generally in agreement for log$_{10}$(M$_{\text{\hi}}$[\msun]) < 10.1. The s50nofb case consistently shows the most extreme trends, highlighting the crucial role of feedback in shaping galaxy structure. 
    }
    \label{fig:Rturn}
\end{figure}

The panels in Fig.~\ref{fig:All_HI_radial_profiles} show the mean \hi\ radial profiles for the different galaxy samples of the  different runs, arranged in order of increasing \hi\ mass. It is evident from Fig.~\ref{fig:All_HI_radial_profiles} that across all mass bins, the mean radial profiles of the various runs differ somewhat at small radii ($R/R_\text{\hi} \lesssim 0.5$), but tend to converge at larger radii, $R/R_\text{\hi} > 0.5$. Specifically, the outer profiles are well described by exponential functions with similar rates of decline whereas the inner profiles tend to be more constant with possible depressions at the center. For the outer segments, we extract the fitted gradients and plot them as a function of \hi\ mass in Fig.~\ref{fig:Rturn}. While the slope of the outer profiles gradually decreases with increasing \hi\ mass, it remains highly consistent across all physics runs at fixed \hi\ mass. The outer \hi\ distributions of \simba\ galaxies are therefore quantifiably self-similar, which we interpret as the underlying reason why the \hi MSR remains robust across different feedback implementations.\\

In contrast to results from the EAGLE simulation \citep{crain2015eagle,schaye2015eagle}—where strong feedback mechanisms led to central \hi\ depressions and notable deviations in the \hi MSR \citep{bahe2016distribution}—our findings reveal no such anomalies, even when individual feedback modes are disabled. Both the \hi MSRs and the \hi\ radial profiles remain closely aligned across all scenarios , with only minor variations, mostly in terms of scatter. This consistency strengthens the case that galaxies adhere to the observed \hi MSR irrespective of feedback strength. In semi-analytic models with resolved disc structure – \cite{fu2013star}, the universal shape of outer gas discs has been attributed to assumptions of exponentially declining gas infall and the "inside-out" formation of discs \citep{wang2014observational}. Feedback is expected to modulate this growth by regulating the availability of cold gas, thereby influencing the pace and nature of inside-out evolution  \citep{baker2025core}. As many galaxy formation models adopt an exponential form for the \hi\ surface density profile, they naturally reproduce similar mass–size scalings \citep{stevens2019origin}.\\

The persistence of the \hi MSR across the various \simba\ runs may stem from the standard methodology used to define \hi\ disc extents. Typically, a surface density threshold of 1 \msunpc\ is adopted, corresponding to galactocentric radii of several kiloparsecs. At these scales, the gas lies beyond the immediate influence of AGN-driven feedback from the galactic centre, thereby preserving the integrity of the \hi MSR. Moreover, this contour generally encloses regions deep within the galaxy’s gravitational potential, where \hi\ is relatively stable against environmental processes such as ram-pressure stripping. However, at lower surface density thresholds (e.g. 0.1 or 0.01 \msunpc), where the \hi\ is more weakly bound, external perturbations are expected to play a more prominent role. Investigating potential deviations in the \hi MSR within these low-density regimes is deferred to future work.

\section{Conclusion}

We investigate the potential impact of different feedback mechanisms on the \hi\ size–mass relation (\hi SMR) using the \simba\ suite of cosmological hydrodynamical simulations at $z = 0$. These simulations incorporate both stellar and AGN-driven feedback processes in the 50 Mpc$^{3}$ box. Our analysis considers four feedback variants: the \textsc{s50nox} run, which excludes only the AGN X-ray feedback mode; \textsc{s50nojet}, which disables both AGN jets and X-ray heating; \textsc{s50noAGN}, which removes all AGN feedback modes, retaining only stellar feedback; and \textsc{s50nofb}, where all feedback processes are turned off.\\

We find that the \hi MSR remains remarkably consistent across all feedback implementations, indicating that the relation for \simba\ galaxies is not strongly regulated by the presence or absence of AGN or stellar feedback. To explore the origin of this invariance, we analyse the azimuthally averaged radial \hi\ surface density profiles for galaxies in each simulation. While mild variations are present at small radii, the profiles converge at larger radii to a common, self-similar exponential form. The outer profile slopes show minimal variation at fixed \hi\ mass across all feedback scenarios, reinforcing the conclusion that the structural uniformity of outer \hi\ discs underlies the observed stability of the \hi MSR.\\

\section*{Data Availability}


Measurement quantities will be made available upon reasonable request.



\bibliographystyle{mnras}







\bsp	
\label{lastpage}
\end{document}